# STUDY INFLUENCE OF DOPING ON TEMPERATURE DEPENDENCE OF DIFFUSION RATE IN ORGANIC MOLECULAR CRYSTALS

## M. A. Korshunov

Kirensky Institute of Physics, Siberian Division, Russian Academy of Sciences, Krasnoyarsk, 660036, Russia (e-mail: mkor@iph.krasn.ru)

**Abstract** It is experimentally shown, that doping an organic molecular crystal reduces diffusion rate of impurity. Also reduces its influence on a modification of temperature. Examinations are lead on a crystal paradibromobenzene doping by paradiclorobenzene. Diffusing impurity was parabromochlorobenzene.

Usually at an increase of temperature diffusion rate increases, it influences physical properties of substance. And for of some practical applications it is undesirable. There is a problem how to reduce agency of an increase of temperature on diffusion rate. Diffusion rate is related not only to a direction of migration of a molecule, but also with altitude of potential barrier parting a migrating molecule from vacancy. Usually with an increase of temperature the distance between molecules is incremented, and the altitude of a barrier decreases, that affects magnification of diffusion rate. Therefore it is necessary to discover requirements at which the altitude of a barrier with an increase of temperature does not decrease. It also was a problem of the yielded operation.

In-process [1] on the basis of calculations it has been shown, that in the mixed molecular crystals magnitude of potential barrier at migration of molecules depends on a modification of temperature, than in their components less.

For check of these calculations experimental researches have been lead. Has been chosen paradibromobenzene doping by paradiclorobenzene. By diffusing impurity has been selected parabromochlorobenzene. These substances have been chosen, as they are isomorphous and crystallize in one space group $P2_1/a$ with two molecules in lattice cell and form mix-crystals of substitution at any concentrations of builders [2]. Single crystals pellucid, that allows to gain qualitative Raman spectrums. Thus spectrums of the lattice oscillations are similar to spectrums of builders. Single crystals of solid solutions of studied substances have been grown and Raman spectra in the field of the lattice and intramolecular oscillations (up to 400 $cm^{-1}$) are gained. In-process concentration of builders was measured in molar unities. For examination of diffusion the method offered in-process [3] allowing discovering concentration of builders in a specimen on spectrums was used.

First spectrums of standard samples with the given relationship p-$C_6H_4Br_2$ and p-$C_6H_4Cl_2$ have been gained, and also p-$C_6H_4BrCl$ and graphs of association of concentration of builders from frequency and intensity of the valence intramolecular oscillations are builted.

Then the single crystal paradibromobenzene a doping by paradiclorobenzene on method Bridgman with the given concentration has been grown. From the grown single crystal it was cut out параллепипед, one of which basils was perpendicular to the chosen crystallographic direction, on this basil the stratum parabromochlorobenzene by thickness ~0.03 see was plotted After that the single crystal was slited on three parts to collaterally chosen crystallographic direction. Each specimen was exposed to annealing during 360 hr.

For each specimen the annealing temperature ($75^0C$ and $70^0C$) has been chosen. It has allowed defining association of diffusion rate (parabromochlorobenzene) from temperature at the yielded concentration. To expel surface effects and to explore diffusion in depth of a specimen from one of basils there was a cut layer of a specimen thickness 0.1cm parallel a direction of diffusion. The specimen found room in the installation intended for examination of diffusion using a method of a Raman scattering of light [3]. Examinations were spent apart 30 from edge of a crystal where has been plotted parabromochlorobenzene.

Concentration parabromochlorobenzene with (discovered on a method resulted in-process [3]) makes $C^0_{75}$ =34.0±2.0 mol %, $C^0_{70}$=13.0±2.0 mol % where the subscript designates annealing temperature of a specimen, and a superscript concentration doping. At magnification doping it has been gained $C^3_{75}$ =30.0±2.0 mol %, $C^3_{70}$=14.0±2.0 mol %. At doping paradiclorobenzene up to 5 % it is gained $C^5_{75}$ =27.0±2.0 mol %, $C^5_{70}$=13.0±2.0 mol %. As we see at magnification of concentration doping diffusion rate decreases. It is manifested in decrease of concentration of impurity (parabromochlorobenzene) measured on one distance and at one temperature $75^0C$. As convergence of values of concentration of impurity for different annealing temperature is noted at magnification doping. It is possible to explain this behavior to that the altitude of potential barrier depends on temperature less, than in недопированных specimens. Though it, apparently, should depend on that what substance are used for doping. That demands the further examination.

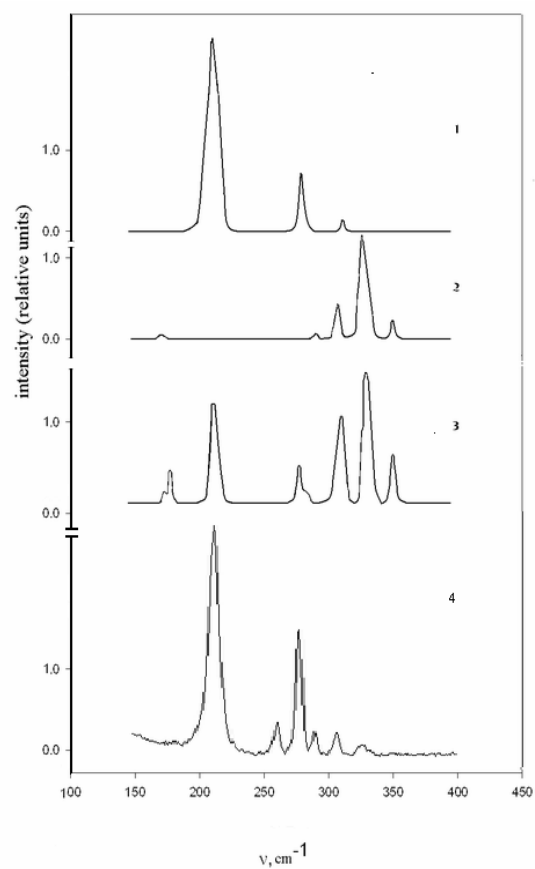

Fig.1. Spectrums of intramolecular oscillations (in the field of = 150 - 400 $cm^{-1}$) paradibromobenzenea (1), paradiclorobenzene (2) parabromochlorobenzene (3) and the studied mixed crystal at concentration of builders (4) (p-$C_6H_4Br_2$ 80-mol % and p-$C_6H_4Cl_2$ 10-mol % and also p-$C_6H_4BrCl$ 10-mol %).